\documentstyle[12pt]{article}

\catcode`\@=11
\long\def\@makefntext#1{
\protect\noindent \hbox to 3.2pt {\hskip-.9pt
$^{{\ninerm\@thefnmark}}$\hfil}#1\hfill}		

\def\@makefnmark{\hbox to 0pt{$^{\@thefnmark}$\hss}}  

\def\ps@myheadings{\let\@mkboth\@gobbletwo
\def\@oddhead{\hbox{}
\rightmark\hfil\ninerm\thepage}
\def\@oddfoot{}\def\@evenhead{\ninerm\thepage\hfil
\leftmark\hbox{}}\def\@evenfoot{}
\def\sectionmark##1{}\def\subsectionmark##1{}}

\renewcommand{\thefootnote}{\fnsymbol{footnote}}

\newcounter{sectionc}\newcounter{subsectionc}\newcounter{subsubsectionc}
\renewcommand{\section}[1] {\vspace*{0.6cm}\addtocounter{sectionc}{1}
\setcounter{subsectionc}{0}\setcounter{subsubsectionc}{0}\noindent
	{\normalsize\bf\thesectionc. #1}\par\vspace*{0.4cm}}
\renewcommand{\subsection}[1] {\vspace*{0.6cm}\addtocounter{subsectionc}{1}
	\setcounter{subsubsectionc}{0}\noindent
	{\normalsize\it\thesectionc.\thesubsectionc. #1}\par\vspace*{0.4cm}}
\renewcommand{\subsubsection}[1]
{\vspace*{0.6cm}\addtocounter{subsubsectionc}{1}
	\noindent
{\normalsize\rm\thesectionc.\thesubsectionc.\thesubsubsectionc.
	#1}\par\vspace*{0.4cm}}

\newcounter{appendixc}
\newcounter{subappendixc}[appendixc]
\newcounter{subsubappendixc}[subappendixc]

\renewcommand{\appendix}[1] {\vspace*{0.6cm}
        \refstepcounter{appendixc}
        \setcounter{figure}{0}
        \setcounter{table}{0}
        \setcounter{equation}{0}
        \renewcommand{\thefigure}{\Alph{appendixc}.\arabic{figure}}
        \renewcommand{\thetable}{\Alph{appendixc}.\arabic{table}}
        \renewcommand{\theappendixc}{\Alph{appendixc}}
        \renewcommand{\theequation}{\Alph{appendixc}.\arabic{equation}}
        \noindent{\bf Appendix \theappendixc #1}\par\vspace*{0.4cm}}

\def\abstracts#1{{

\centering{\begin{minipage}{12.2truecm}\footnotesize\baselineskip=12pt\noindent
	\centerline{\footnotesize ABSTRACT}\vspace*{0.3cm}
	\parindent=0pt #1
	\end{minipage}}\par}}

\renewenvironment{thebibliography}[1]
	{\begin{list}{\arabic{enumi}.}
	{\usecounter{enumi}\setlength{\parsep}{0pt}
\setlength{\leftmargin 1.25cm}{\rightmargin 0pt}
	 \setlength{\itemsep}{0pt} \settowidth
	{\labelwidth}{#1.}\sloppy}}{\end{list}}

\newcounter{itemlistc}
\newcounter{romanlistc}
\newcounter{alphlistc}
\newcounter{arabiclistc}

\newcommand{\fcaption}[1]{
        \refstepcounter{figure}
        \setbox\@tempboxa = \hbox{\footnotesize Fig.~\thefigure. #1}
        \ifdim \wd\@tempboxa > 6in
           {\begin{center}
        \parbox{6in}{\footnotesize\baselineskip=12pt Fig.~\thefigure. #1}
            \end{center}}
        \else
             {\begin{center}
             {\footnotesize Fig.~\thefigure. #1}
              \end{center}}
        \fi}

\newcommand{\tcaption}[1]{
        \refstepcounter{table}
        \setbox\@tempboxa = \hbox{\footnotesize Table~\thetable. #1}
        \ifdim \wd\@tempboxa > 6in
           {\begin{center}
        \parbox{6in}{\footnotesize\baselineskip=12pt Table~\thetable. #1}
            \end{center}}
        \else
             {\begin{center}
             {\footnotesize Table~\thetable. #1}
              \end{center}}
        \fi}

\def\@citex[#1]#2{\if@filesw\immediate\write\@auxout
	{\string\citation{#2}}\fi
\def\@citea{}\@cite{\@for\@citeb:=#2\do
	{\@citea\def\@citea{,}\@ifundefined
	{b@\@citeb}{{\bf ?}\@warning
	{Citation `\@citeb' on page \thepage \space undefined}}
	{\csname b@\@citeb\endcsname}}}{#1}}

\newif\if@cghi
\def\cite{\@cghitrue\@ifnextchar [{\@tempswatrue
	\@citex}{\@tempswafalse\@citex[]}}
\def\citelow{\@cghifalse\@ifnextchar [{\@tempswatrue
	\@citex}{\@tempswafalse\@citex[]}}
\def\@cite#1#2{{$\null^{#1}$\if@tempswa\typeout
	{IJCGA warning: optional citation argument
	ignored: `#2'} \fi}}

 1
 1
 1

\font\ninerm=cmr9

\begin{document}
\begin{flushright}
UH-511-821-95  \\
March 1995
\end{flushright}

\newcommand{\st}{\scriptstyle}
\newcommand{\sst}{\scriptscriptstyle}
\newcommand{\mco}{\multicolumn}
\newcommand{\epp}{\epsilon^{\prime}}
\newcommand{\vep}{\varepsilon}
\newcommand{\ra}{\rightarrow}
\newcommand{\ppg}{\pi^+\pi^-\gamma}
\newcommand{\vp}{{\bf p}}
\newcommand{\ko}{K^0}
\newcommand{\kb}{\bar{K^0}}
\newcommand{\al}{\alpha}
\newcommand{\ab}{\bar{\alpha}}
\def\be{\begin{equation}}
\def\ee{\end{equation}}
\def\bea{\begin{eqnarray}}
\def\eea{\end{eqnarray}}
\def\CPbar{\hbox{{\rm CP}\hskip-1.80em{/}}}

\centerline{\normalsize\bf NEUTRINO OSCILLATIONS WITH
BEAMS FROM AGN'S AND GRB'S\footnote{Presented at the {\it Fourth
International Conference on Physics
Beyond Standard Model}, Lake Tahoe, CA, December 13-18, 1994.}}
\baselineskip=22pt
\baselineskip=16pt

\centerline{\footnotesize SANDIP PAKVASA}
\baselineskip=13pt
\centerline{\footnotesize\it Department of Physics and Astronomy,
University of Hawaii,}
\baselineskip=12pt
\centerline{\footnotesize\it Honolulu, HI  96822, USA}
\vspace*{0.3cm}
\baselineskip=13pt

\vspace*{0.9cm}
\abstracts{I discuss how a 1 KM3 neutrino detector can be used to study
$\nu_\tau$ oscillations at PeV energies with neutrinos from AGN's and to
study neutrinos from GRB's.}

\normalsize\baselineskip=15pt
\setcounter{footnote}{0}
\renewcommand{\thefootnote}{\alph{footnote}}
\section{Introduction}
I would like to discuss two topics: (i) the observation of $\tau's$ from
$\nu_\tau's$ produced in oscillations of $\nu_\mu's$ from Active Galactic
Nuclei with energies of a few PeV and (ii) possible observation of
$\nu_\mu's$ and $\nu_e's$ from Gamma Ray Bursters with energies ranging from
MeV to TeV.  Details and a complete set of references are to be found in
References [1] and [2].

The main assumption I make is that a next generation DUMAND-like water
Cerenkov under-$H_2O$ array of dimensions of order of $1 km^3$ will be
available
at some future date to detect neutrino interactions.

\section{Neutrinos from Active Galactic Nuclei}

	For AGN's the expectations are that they emit high energy $\nu's$;
the total flux overtakes atmospheric $\nu$-flux by $E_\nu \sim 0 (TeV)$ and the
most likely flavor mix is $\nu_\mu: \nu_e : \nu_\tau \approx 2:1:0.$  This
is my second major assumption.

\subsection{$\nu_\tau$ Signature}
For a $\nu_\tau$ of energy above 2 PeV there is a characteristic ``double
bang'' signature.   When $\nu_\tau$ interacts via charged current there is a
hadronic shower (of energy $E_1)$ with about $10^{11}$ photons emitted; then
the
$\tau$
travels about 90m (for $E_\tau \sim 1.8 PeV)$ and when it decays (either to
e's or hadrons with 80 \% probability) there is again a cascade (of energy
$E_2)$
with $2.10^{11}$ photons emitted in Cerenkov light.  The $\tau$ track is
minimum
ionising and may emit $10^6-10^7$ photons; even if it is not resolvable, one
can connect the two showers by speed of light and reconstruct the event.

The backgrounds (after appropriate cuts) are very small.  Hence such
``double bang'' events represent either $\nu_\mu \rightarrow \nu_e$ (or
$\nu_e \rightarrow \nu_\tau)$ oscillations or $\nu_\tau$-emission at the
source and in any case are extremely interesting.  For signal events due to
$\nu_\tau$, one expects $E_2/E_1 > 2$ on the average, and hence a cut of
$E_2/E_1 > 1$
removes many backgrounds; another cut
on the distance D between the two bangs of $D > 50m$ eliminates most of the
punch-thru backgrounds.

	Extra bonuses from observing these double bang events are:  (i) use
of the zenith angle distribution to measure $\sigma_\nu$ via attenuation and
(ii) use of the enormous light collection and good timing to get good vertex
resolution and determine $\nu_\tau$ direction to within one degree.

\subsection{Expected Flavor Mixes}
Most models of $\nu$-emission in AGN's correspond to tenuous beam dumps with
little absorption and $\nu's$ come from $\pi$ (and K) decay.  Frequently
$\gamma p \rightarrow  \Delta$ is a dominant process.  In these scenarios we
expect at production
\begin{eqnarray*}
\nu_\mu: \ \nu_e: \nu_\tau \approx 2 : 1 : 0
\end{eqnarray*}

For example, in the Protheroe-Szabo model, they find $\nu_\mu :\nu_e
\approx 1.75:1$ and 10\% of $\nu's$ come from pp interactions.  Some
fraction of pp collisions will contribute to prompt $\nu's$ (including
$\nu_\tau's)$ via production of c and b. In the prompt $\nu's$ the flavor mix
is
\begin{eqnarray*}
\nu_\mu: \nu_e:  \nu_\tau = 1 : 1 : p
\end{eqnarray*}
where $p$ can be crudely estimated to be about $0.07$ to 0.1. Since the
prompt $\nu's$ themselves are expected to be only 10\% of total the modified
flavor mix is
\begin{eqnarray*}
\nu_\mu:  \nu_e : \nu_\tau \approx 1: 0.6 : 0.01
\end{eqnarray*}
and contains less than 1\% of $\nu_\tau's$.

\subsection{Rates}

To estimate event rates we make the following assumptions:
(i) assume the fluxes of Protheroe-Szabo model
(ii) integrate over all AGN's
(iii) assume an initial flavor mix of $\nu_\mu : \nu_e \nu_\tau \approx
2:1:0$ and $\nu_\mu \rightarrow \nu_\tau$ conversion such that on arrival
the flavor mix is $\nu_\mu : \nu_e : \nu_\tau = 1: 1 : 1$
(iv) a 1km$^3$ water \^{c} detector with 100 \% detection efficiency.
Then we expect 1000 $\nu_\tau$ ``double bang'' events and about 1800
showering events $(\nu_e CC$ and $\nu_\alpha$ NC) per year.

\subsection{Oscillations}

The neutrino flavor mix can be easily determined from the event
classification of the data.  The double bang events determine $\nu_\tau +
\bar{\nu}_\tau$ flux; the upcoming muons determine $\nu_\mu + \bar{\nu}_\mu$
flux; the cascade events (single bang) determine a
combination of $(\nu_e + \bar{\nu}_e), (\nu_\mu + \bar{\nu}_\mu)$ and
$(\nu_\tau
+
\bar{\nu}_\tau)$ fluxes; and Glashow Resonance (W) events determine
$\bar{\nu}_e$ flux (at $E_\nu = 6.4 PeV)$.

The sensitivity to oscillation parameters depends on several factors.  If
individual AGN's can be seen in $\nu_\tau's$ (say to 100 mpc) then $\delta
m^2$ upto $\geq 10^{-16} ev^2$ can be probed.  The mixing angle sensitivity is
limited by
$\sin^2 2 \theta > 0.01$ due to some initial $\nu_\tau$ present and in
practice probably closer to 0.05 to 0.1.

To proceed further let us assume:  (i) initial fluxes are $\nu_\mu : \nu_e :
\nu_\tau \approx 2:1:0$; (ii) \# $\nu$ = \# $\bar{\nu}$ (although this is not
essential); (iii) all $\delta m^2 >> 10^{-16} eV^2,$
i.e.
$<sin^2~(\delta m^2 L/4
E) > \approx 1/2$; (iv) matter effects negligible at production (e.g. $N_{e-}
= N_{e +})$ and no significant matter effects en-route (this is valid for
$\delta m^2$ of current interest $\sim 10^{-2} - 10^{-6} ev^2);$ (v)
atmospheric $\nu$-anomaly caused by $\nu_\mu - \nu_\tau$ oscillations with
$\delta m^2 \sim 10^{-2} ev^2$ and $\sin^2 2 \theta \geq 0.6$.  In this case we
expect
$\nu_\mu : \nu_e : \nu_\tau \approx 1: 1: 1$ at earth.  If on the other
hand, the atmospheric anomaly is due to $\nu_\mu-\nu_e$ mixing the flavor mix
on arrival is 1:1:0.  These two cases are easy to distinguish with the annual
rate of
double bang events varying from 1000 to zero.  We have also considered 3
neutrino mixing with solar neutrinos accounted for by either MSW or LWO
oscillations and found a large $\nu_\tau$ flux is always found; and the
various solutions can be distinguished by the different $\nu_\mu \nu_e,
\nu_\tau$ fluxes observed.  As two limiting cases of interest:  (i) a pure
prompt flux 1:1:01 becomes 1:1.6 :0.7 with $\nu_\mu \rightarrow
\nu_\tau$ conversion and is very distinct and (ii) an initial universal flux
1:1:1 remains universal!

\subsection{Backgrounds and Summary}

We have considered several possible sources of backgrounds which fake
double bang signatures. The most serious appears to be a $\nu_\mu
\rightarrow \mu$ charged current event where the $\mu$ travels about 100 m
without much radiation and then deposits the bulk of its energy in a
catastrophic bremstrahlung.  This would have all the characteristics of a
genuine $\nu_\tau$ event.  We estimate the fraction of such events to be
about $(m_e/m_\mu)^2 (100 m/R_\mu)(\Delta E/E) \sim 3.10^{-3}$ and
seems reassuringly small.

At the hadronic vertex, the sources of background are:  (i) $\nu_e + N
\rightarrow e + D_s$ produced diffractively with $D_s \rightarrow \tau \nu$;
and $E_2/ E_1$ can be of 0(1) to fake the $\nu_\tau$ signal provided $D_s$
decays quickly.  The rate is expected to be of order $3.10^{-4}$ of cc events;
(ii) $\nu_{\alpha} + N \rightarrow \nu_\alpha + D_s/B$  again with $D_s$ or
B decaying into $\tau$ within 10m and $\tau$ traveling 100 m.  In these
events we expect $E_2/E_1 < 1$ and again the rate is small of order $\sim
10^{-3}$. Other backgrounds such as coincident downgoing $\mu's$ showering is
expected to be small.  Hence, that after the cuts such as $E_2/E_1 > 1
$ and $D > 50 m,$ the backgrounds are rather small.

We conclude that given AGN $\nu$-sources, it is possible to see $\nu_\tau
\rightarrow \tau$ events in a 1 km3 array unambigously.  One can measure
$\nu_\tau$ mixing angles $(sin^2 2 \theta > 0.1)$ and $\delta m^2 (>
10^{-16} eV^2$) and/or determine the presence of $\nu_\tau$ in the initial
beam.
If $\nu_\tau$ mixing were known from terrestrial experiments one would learn
about the nature of AGN $\nu$-emission processes.  {\it We feel that this
already justifies the construction and deployment of a $1km^3$ array.}

\section{Gamma Ray Bursters}

The gamma ray bursters remain rather enigmatic; but the current indications
are that they are probably at cosmological distances with very high
luminosities.  Their luminosities seem to be in a narrow range and they may be
almost standard candles.  The photon energies occasionally reach GeV and
the spectrum may be power law $(\sim 1/E^2)$.  Perhaps the gamma rays come
from $\pi^0$ decay, in which case there should be $\nu_\mu's$ from $\pi^\pm$
decay.  In models like those of Paczynski-Xu this is indeed the case.  We
assume this is true and take as nominal values $\phi_\gamma \sim 1 per cm^2$
at 1 MeV, $\phi_\gamma \sim 1/E^2$ and let $\phi_\nu \sim \eta \phi_\gamma$.
In
some models $\eta$ can be much larger than one.  The flavor mix should be
$\nu_\mu: \nu_e \sim 2:1$ as in atmospheric neutrinos and modified by
osscillations.  In principle one can be sensitive to $\delta m^2$ to
$10^{-19} eV^2$.  Many other neutrino properties can be probed to new levels.
An exciting new twist is the ability to measure Hubble red shift in
neutrinos, providing a first non-electromagnetic test of the expansion.

Typical rates in neutrino telescopes such as DUMAND, AMANDA etc. are $\sim
1.5.10^{-2} \eta$ correlated $\mu's$ per year for $E_{th}$ of 20 GeV.  For
the 1km3 detector the number is $1.5 \eta$ correlated $\mu's$ per year
and $10^{-2} \eta \ \mu's$ per burst.
Considering $\eta$ can be as high as 100 these are non-negligible.  For the
brightest 0.1\% of GRB's, one may witness a burst of $100 \mu's$ once a
year.

Any detection of GRB'S in neutrinos would be fantastic.  But to make this
happen, we need $\nu$-Telescopes with largest possible area (for high
energies), volume (for low energies) and lowest possible threshold (to
increase rate as well as to do flavor analysis etc).

I believe we have given a strong rationale for building a $1 km^3$
neutrino detector and shown that there is very exciting physics and
astrophysics out there waiting to be discovered.

\section{Acknowledgments}

The contents of this talk are based on enjoyable collaboration with Tom
Weiler, Walter Simmons and especially John Learned.  This work was
supported in part by U.S.D.O.E. under grant DE-FG 03-94ER40833.

%
\normalsize

\section{References}




\begin{thebibliography}{9}
\bibitem{1}  J. G. Learned and S. Pakvasa, UH-511-799-94, AstroParticle
Physics Journal (in press), hep-ph/9405296.
\bibitem{2} T. J. Weiler, W. A. Simmons, S. Pakvasa and J. G. Learned,
UH-511-801-94, hep-ph/9411432.

\end{thebibliography}
\end{document}